\documentclass[a4paper]{jpconf}
\usepackage[symbol]{footmisc}
\usepackage{graphicx}
\usepackage{soul,color}
\usepackage{physics}
\usepackage{lineno}
\usepackage[utf8]{inputenc}
\usepackage[english]{babel}
\usepackage{fancyhdr}
\usepackage{lastpage}
\makeatletter %% <- make @ usable in command sequences
\newcount\SOUL@minus
\makeatother  %% <- revert @
\begin{document}
\renewcommand{\thefootnote}{\fnsymbol{footnote}}

\newcommand{\ttH}{\ensuremath{t\bar{t}H}\,}
\newcommand{\Hmumu}{\ensuremath{H\rightarrow\mu^{+}\mu^{-}}\,}
\newcommand{\Hgamgam}{\ensuremath{H\rightarrow\gamma\gamma}\,}

%{\LARGE\textbf{CONFIDENTIAL DRAFT, PLEASE DO NOT DISTRIBUTE}}
%\vspace{1cm}

%Publication Draft Version 2, \today 
%\vspace{1cm}

%\newline Edited by Sau Lan Wu and Chen Zhou, University of Wisconsin 
%(assisted by Wen Guan)

%\vspace{1cm}

%\newline Intended to be submitted to Nature XXX.

\title{Application of Quantum Machine Learning using the Quantum Variational Classifier Method to High Energy Physics Analysis at the LHC on IBM Quantum Computer Simulator and Hardware with 10 qubits}

\author{
%\newline $\,$
%\newline 
Sau Lan Wu\textsuperscript{1*}, Jay Chan\textsuperscript{1}, Wen Guan\textsuperscript{1}, Shaojun Sun\textsuperscript{1}, Alex Wang\textsuperscript{1}, Chen Zhou\textsuperscript{1}\\
%\newline 
Miron Livny\textsuperscript{2}\\
%\newline 
Federico Carminati\textsuperscript{3}, Alberto Di Meglio\textsuperscript{3}\\
%\newline Panagiotis Barkoutsos\textsuperscript{4}, Ivano Tavernelli\textsuperscript{4}, Stefan Woerner\textsuperscript{4} 
%\newline 
Andy C. Y. Li\textsuperscript{4}, Joseph Lykken\textsuperscript{4}, Panagiotis Spentzouris\textsuperscript{4}\\
%\newline Jennifer Glick\textsuperscript{6}
%\newline 
Samuel Yen-Chi Chen\textsuperscript{5}, Shinjae Yoo\textsuperscript{5}\\
%\newline 
Tzu-Chieh Wei\textsuperscript{6}\\
}

\vspace*{0.5pc}%
\begin{indented} 
\item[]\noindent\normalsize 
\textsuperscript{1}Department of Physics, University of Wisconsin, Madison WI, USA \\ %
\textsuperscript{2}Department of Computer Sciences, University of Wisconsin, Madison WI, USA  \\ %
\textsuperscript{3}CERN Quantum Technology Initiative, IT Department, CERN, Geneva, Switzerland \\ %
%\textsuperscript{4}IBM Research - Zurich, Rüschlikon, Switzerland \\ %
\textsuperscript{4}Quantum Institute, Fermi National Accelerator Laboratory, Batavia, IL, USA \\ %
%\textsuperscript{6}IBM T.J. Watson Research Center, Yorktown Heights, NY, USA \\ %
\textsuperscript{5}Computational Science Initiative, Brookhaven National Laboratory, Upton, NY, USA \\ %
\textsuperscript{6}C.N. Yang Institute for Theoretical Physics, State University of New York at Stony Brook, Stony Brook, NY, USA \\ %
\end{indented}

\clearpage
\begin{abstract}
%The ambitious HL-LHC program will require enormous computing resources in the next two decades. 
%A burning question is whether quantum computer can solve the ever growing demand of computing resources in High Energy Physics in general and physics at the LHC in particular. 
%The experimental programs at the LHC revolve around one major objective: 
One of the major objectives of the experimental programs at the LHC is
the discovery of new physics. This requires the identification of rare signals in immense backgrounds. Using machine learning algorithms greatly enhances our ability to achieve this objective. 
With the progress of quantum technologies, quantum machine learning could become a powerful tool for data analysis in high energy physics. 
In this study, using IBM gate-model quantum computing systems, we employ the quantum variational classifier method in two recent LHC flagship physics analyses: \ttH\ (Higgs boson production in association with a top quark pair, probing the Higgs boson couplings to the top quark) and \Hmumu\ (Higgs boson decays to two muons, probing the Higgs boson couplings to second-generation fermions). 
We have obtained early results with 10 qubits on the IBM quantum simulator and the IBM quantum hardware. 
With small training samples of 100 events on the quantum simulator, the quantum variational classifier method performs similarly to classical algorithms such as SVM (support vector machine) and BDT (boosted decision tree), which are often employed in LHC physics analyses. 
On the quantum hardware, the quantum variational classifier method has shown promising discrimination power, comparable to that on the quantum simulator. 
This study demonstrates that quantum machine learning has the ability to differentiate between signal and background in realistic physics datasets.
%In the future, by exploiting the high dimensional quantum feature space expanded by a large number of qubits and by mitigating the impact of quantum hardware noise, quantum computers can possibly achieve advantages in the performance of machine learning classifiers.
We foresee the usage of quantum machine learning in future high-luminosity LHC physics analyses, including measurements of the Higgs boson self-couplings and searches for dark matter. 
%Using IBM Quantum Computer Simulators and Quantum Computer Hardware, we have successfully employed the Quantum Support Vector Machine (QSVM) method in applying quantum machine learning for the \ttH\ (Higgs boson production in association with a top quark pair, probing Higgs couplings to top quarks) analysis and the \Hmumu\ (Higgs boson decays to two muons, probing Higgs couplings to 2nd generation fermions) analysis at the LHC.
\end{abstract}

\clearpage
%\section{INTRODUCTION}
%\vspace{0.2cm}

The discovery of the Higgs boson by the ATLAS and CMS experiments at the Large Hadron Collider (LHC) in 2012~\cite{atlashiggs,cmshiggs} was a major milestone for high energy physics.
Since then, LHC experiments have been using the Higgs boson as a tool to pursue the discovery of new physics. 
%One popular approach is to probe the couplings between the Higgs boson and elementary fermions (i.e. quarks and leptons), a fundamental interaction theorized to be responsible for the fermion masses.
The discovery of new physics requires the identification of rare signals against immense backgrounds. Using machine learning greatly enhances our ability to achieve this objective. 

The intersection between machine learning and quantum computing has been referred to as quantum machine learning, 
and can possibly offer a valuable alternative to classical machine learning by providing more efficient solutions~\cite{qml}.
In 2018, a quantum variational classifier method was experimentally implemented with a quantum circuit of 2 qubits on a superconducting processor and successfully tested on synthetic datasets~\cite{qsvmv}.
%The study have experimentally demonstrated a classifier that exploits a quantum feature space.
This method provides ``tools for exploring the applications of noisy
intermediate-scale quantum computers to machine learning''~\cite{qsvmv}.
With the progress of quantum technologies, quantum machine learning could possibly become a powerful tool for data analysis on real-world datasets such as those seen in high energy physics. 

In this study, we employ the quantum variational classifier method in a \ttH\ (H $\rightarrow$ 2 photons) physics analysis and a \Hmumu\ physics analysis, 
two recent flagship physics analyses at the LHC, using IBM gate-model quantum computers. 
Our goal is to explore and to demonstrate, in a proof of principle experiment, the potential of quantum computers can be a new computational paradigm for big data analysis in high energy physics.
%Our goal is to explore and to demonstrate as a proof of principle that quantum computing can possibly be the new paradigm in the future,
%or at least it can be effectively used for big data analysis.
%In the last six months, we have given three presentations in recent QIS conferences on this topic~\cite{qsvmvpre, qsvmvcern, qsvmveqtc19}. We are extending this experience to other Higgs channels, e.g., Higgs to two muons, double Higgs production etc. and to search for Dark Matter at LHC.
An earlier study in a ggH (H $\rightarrow$ 2 photons) physics analysis using D-wave quantum annealers was performed by A. Mott et al.~\cite{maria}.

\begin{figure}[htb]
\begin{center}
\begin{tabular}{c c c}
    \includegraphics[height=1.2in]{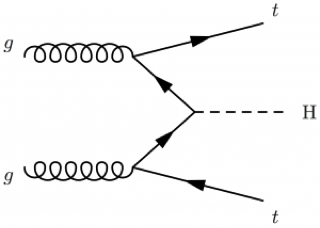} &  \includegraphics[height=1.2in]{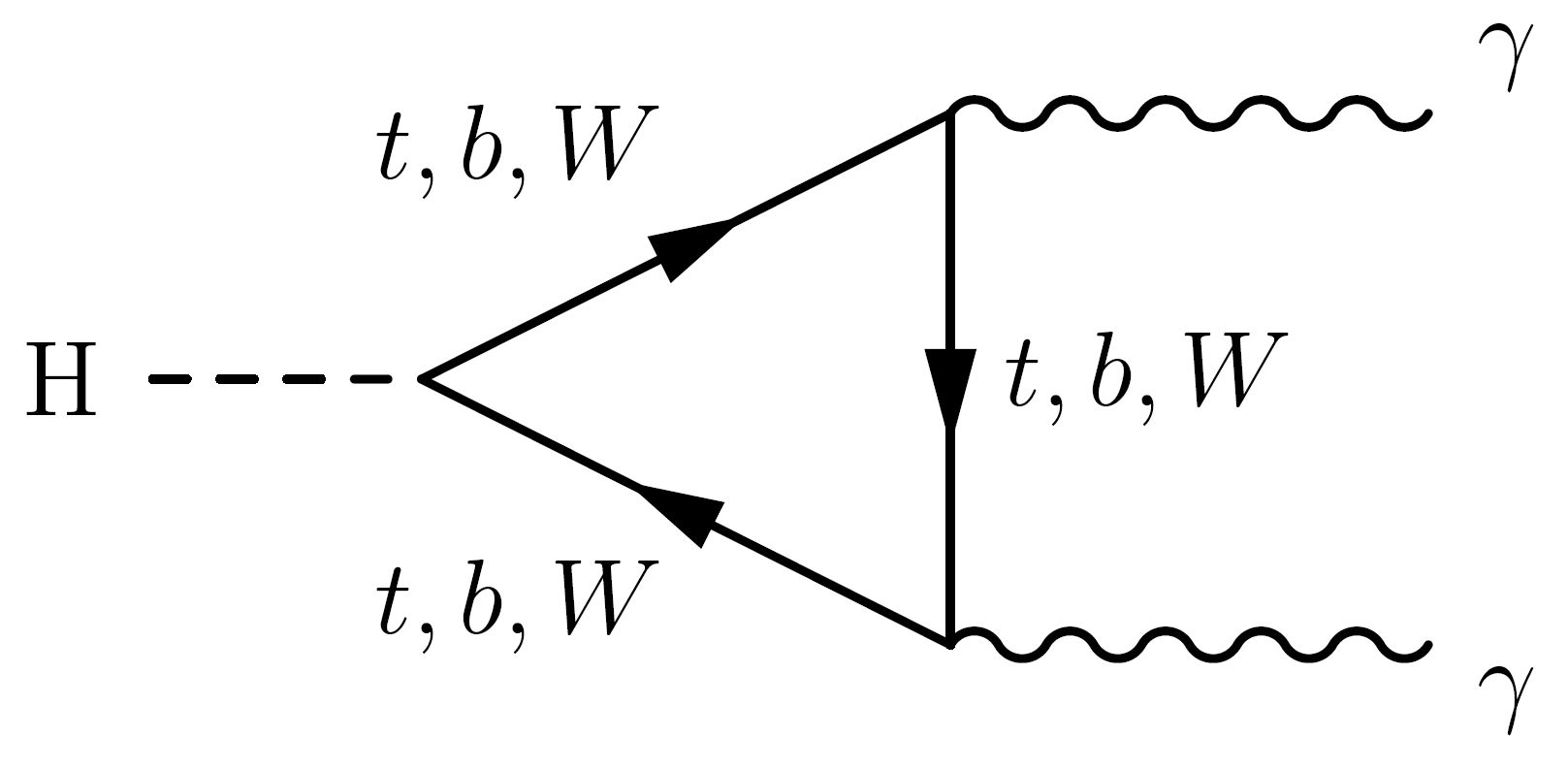} &  \includegraphics[height=1.2in]{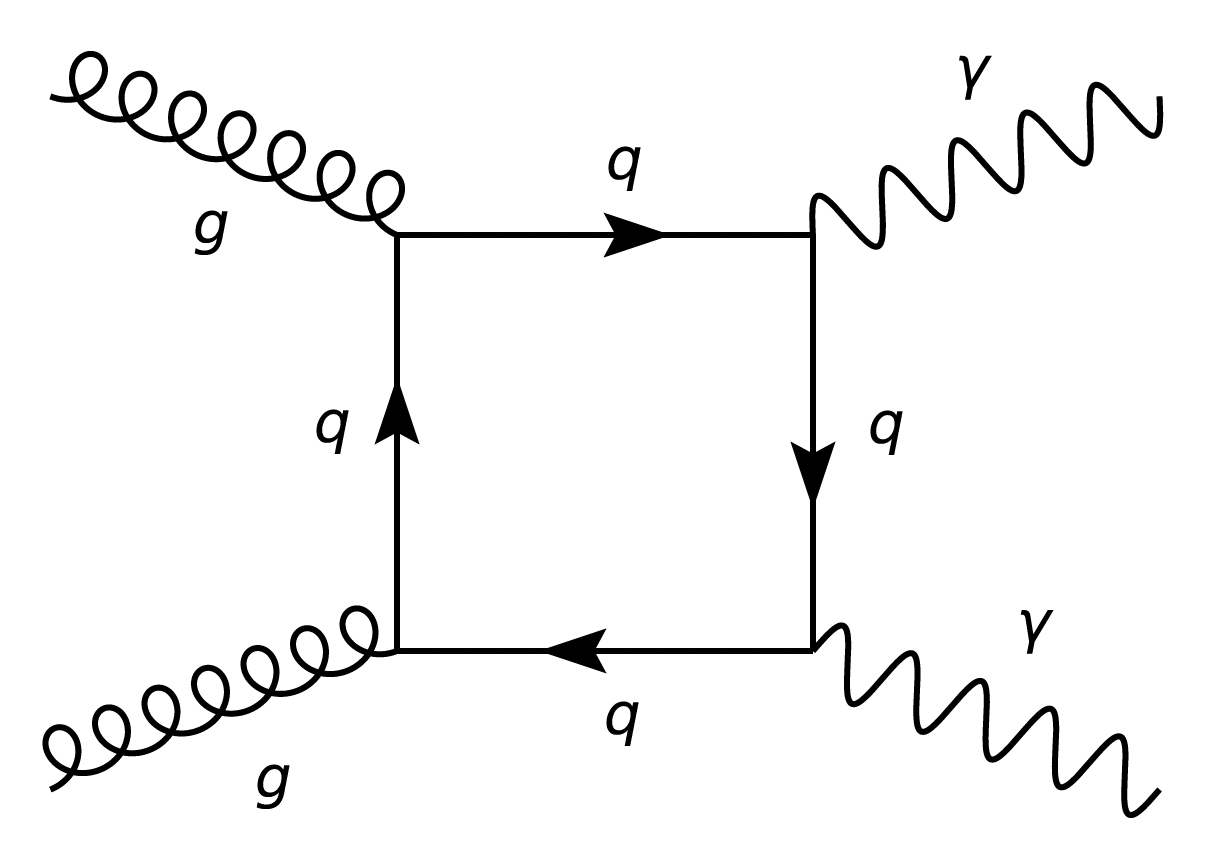} \\
    (a) &  (b) &  (c) \\    
    \includegraphics[height=1.2in]{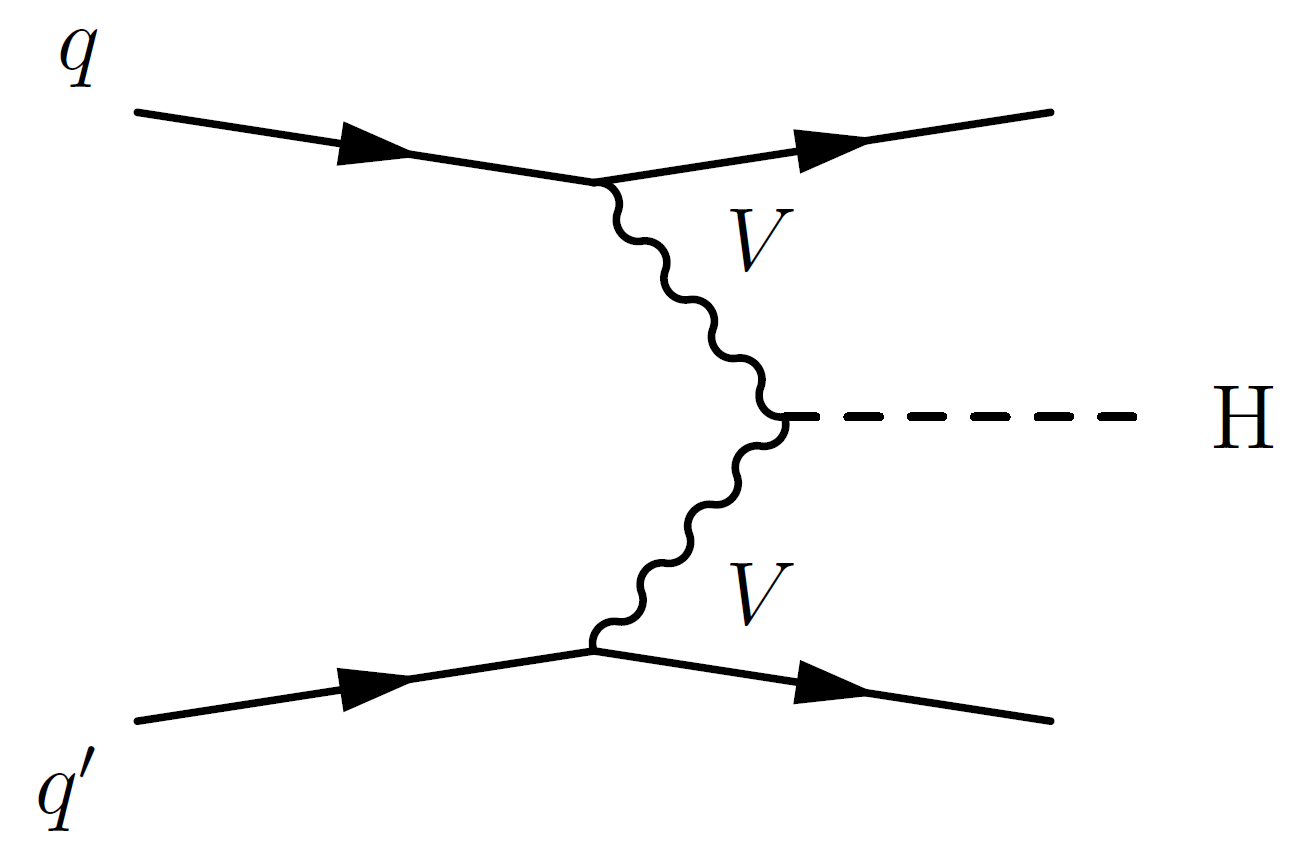} &  \includegraphics[height=1.2in]{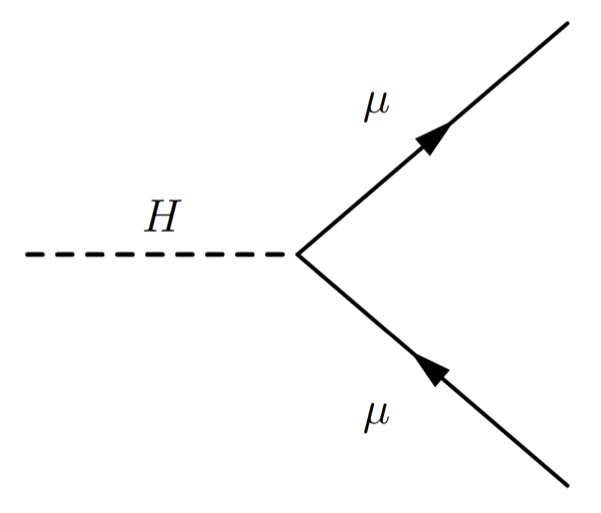} &  \includegraphics[height=1.2in]{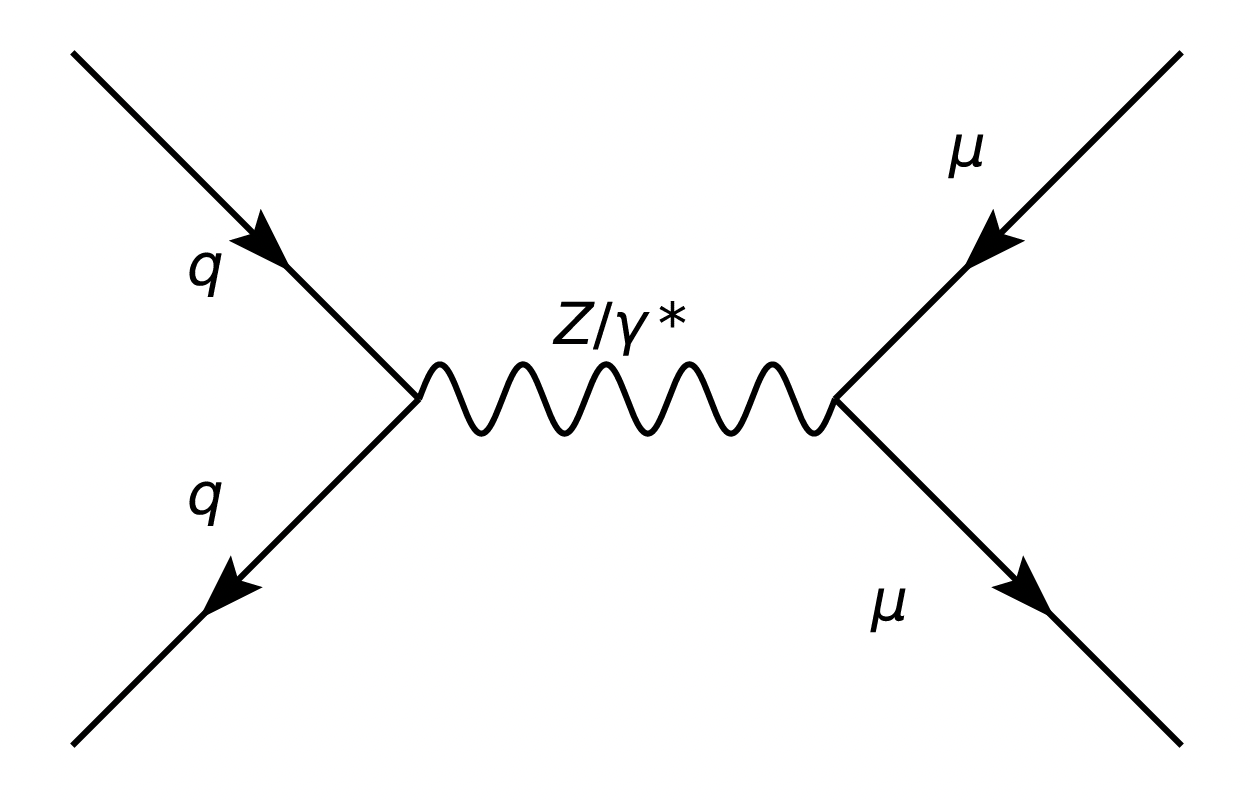} \\
    (d) &  (e) &  (f) \\    
\end{tabular}
\end{center}
\caption{Representative Feynman diagrams for (a) $\ttH$ production, (b) $\Hgamgam$ decay, (c) non-resonant two-photon production, (d) VBF Higgs production, (e) $\Hmumu$ decay,  and (f) $\ensuremath{Z/\gamma^*\rightarrow\mu\mu}$ production. 
In these diagrams, $H$ denotes a Higgs boson, $g$ denotes a gluon, $q$ denotes a quark, $t$ denotes a top quark, $b$ denotes a bottom quark, $\mu$ denotes a muon, $W$ denotes a W boson, $Z$ denotes a Z boson, $V$ denotes a W boson or Z boson, and $\gamma$ denotes a photon.}
\label{fig1}
\end{figure}

%\section{Two recent LHC flagship physics analyses}
\vspace{0.5cm}
\textbf{Two recent LHC flagship physics analyses:}
%Since the discovery of the Higgs boson in 2012, \ttH production is one of the physics processes that the LHC experiments have focused on. 
The observation of \ttH\ production (Higgs boson production in association with a top quark pair) in 2018 by the ATLAS and CMS experiments~\cite{atlastth,cmstth} was a significant milestone for the understanding of fundamental particles and interactions. 
It confirmed the interactions between the Higgs boson and the top quark, which is the heaviest known fundamental particle. 
%\hl{New physics particles could participate in certain productions and decays of the Higgs boson.} 
The measurement of the Higgs-top coupling strength could refine our understanding of the Higgs mechanism and provide important handles to new physics. 
%Its observation was extremely challenging because of the small \ttH\ production cross section at the LHC. 
%Because of the low production rate of \ttH, the data analysis for its observation~\cite{atlastth,cmstth} was extremely challenging.
%As only 1$\%$ of all Higgs bosons are produced in association with two top quarks, its observation was extremely challenging. 
As \ttH\ only accounts for about 1$\%$ of the total Higgs boson production at the LHC, its observation was extremely challenging. 
%Hence it is necessary to exploit all possible Higgs decay channels.
Here we address a channel where the Higgs boson decays into two photons ($\Hgamgam$) and the two top quarks decay into jets. 
%In the ATLAS Collaboration, analysis optimization employing advanced machine learning techniques was successful to obtain the best sensitivity in the measurement of the \ttH (H $\rightarrow$ 2 photons) process~\cite{atlastth}.  
%In this channel, events containing two photons are selected, 
%and split to a Hadronic region (top quarks decay predominantly to hadrons) and a Leptonic region (top quarks decay predominantly to leptons) to target different top quark pair decay topologies.
%In the Hadronic region, a dedicated Boosted Decision Tree (BDT) is trained using the XGBoost package~\cite{xgboost}. %which is currently one of the most advanced BDT algorithms. 
%ttH signal events from simulation and background events from data control regions are both separated into three non-overlapping parts, for BDT training, hyper-parameter tuning, and sensitivity evaluation, respectively. 
To ensure the results are as realistic as possible, we closely follow an analysis strategy similar to that employed by ATLAS~\cite{atlastth}.
Starting from reconstructed events with two photons and at least three jets, we train classifiers to separate the \ttH\ ($\Hgamgam$) signal from the dominant background of this analysis, the non-resonant two-photon production. 
See Figure~$\ref{fig1}$ for representative Feynman diagrams for $\ttH$ production (a), $\Hgamgam$ decay (b), and non-resonant two-photon production (c).
The training is using 23 kinematic variables similar to those in~\cite{atlastth}: 
%using the following input variables: 
the transverse momentum $p_T$, pseudo-rapidity $\eta$ and $b$-tagging status of up to 6 leading jets, the magnitude of the missing transverse momentum, as well as the $p_T$/$m_{\gamma\gamma}$ ($m_{\gamma\gamma}$ denotes invariant mass of the photon pair) and $\eta$ of the two photons.

The searches for \Hmumu\ decay (Higgs boson decay into two muons) at the ATLAS and CMS experiments~\cite{atlashmumu,cmshmumu} have become one of the most important topics in the LHC physics program. 
Although the coupling between the Higgs boson and third-generation fermions (e.g. top quark) has been observed, currently there exist only first indications of the coupling between the Higgs boson and second-generation fermions. 
\Hmumu\ decay is the most promising process by which to observe such a coupling at the LHC.
The strength of the Higgs-muon coupling could be significantly modified by new physics.
With more data in the future, the LHC experiments could establish the Higgs couplings to muons, or exclude the Higgs-muon coupling: both will be an exciting discovery.
In searches for \Hmumu\ decay, the challenge is mainly due to the small \Hmumu\ decay branching ratio of about 0.02$\%$. 
Again following an analysis strategy similar to that used by ATLAS~\cite{atlashmumu}, we divide reconstructed two-muon events into several $n_j$ (jet multiplicity) channels, and focus on the $n_j \geq 2$ channel to target vector boson fusion (VBF) Higgs production, whose signature is two forward jets. 
%a number of boosted decision trees (BDT) are trained in each jet multiplicity channel with the XGBoost package \cite{xgboost}. 
%A "Higgs BDT classifier" is trained in each of the three jet channels with the combined $\mathrm{VBF}+\mathrm{ggF}$ Monte Carlo as the signal sample, and the sideband data from 2015 to 2018 as the background sample. 
We train classifiers to distinguish between the \Hmumu\ signal and the dominant background of this analysis, 
production of a pair of muons through the exchange of a Z boson or a  virtual photon $(\ensuremath{Z/\gamma^*\rightarrow\mu\mu})$. 
See Figure~$\ref{fig1}$ for representative Feynman diagrams for VBF Higgs production (d), $\Hmumu$ decay (e), and $\ensuremath{Z/\gamma^*\rightarrow\mu\mu}$ production (f). 
The training is based on 13 kinematic variables similar to those in ~\cite{atlashmumu}:
the $p_{T}$ and rapidity $Y$ of the two-muon system, the absolute value of the cosine of the lepton decay angle $\cos\theta^*$ in the Collins-Soper frame, the $p_{T}$ and $\eta$ of the two leading jets, the relative azimuthal angle of each jet with respect to the di-muon system, the $p_{T}$, $Y$ and invariant mass of the two-jet system, as well as the relative azimuthal angle between the two-jet system and the two-muon system.

In both the \ttH\ and \Hmumu\ cases, we generate the signal and background events using Madgraph5$\_$aMC@NLO~\cite{madgraph} plus Pythia6~\cite{pythia}. 
The center-of-mass energy of the proton-proton collisions of the generated events is set to 13 TeV (same as the ATLAS publications). 
For each generated event, we simulate the detector response using Delphes~\cite{delphes}. %and evaluate the above-mentioned physics variables.
A Principal Component Analysis (PCA) method~\cite{pca,pca1} is employed for data compression, converting the kinematic variables to a smaller number of PCA variables so that the number of encoded variables matches the number of available qubits (which is 10 in this study). 
%A MinMaxScale is followed after the PCA to standardize the input data ranges between $-\pi$ to $\pi$.
After PCA, the data is transformed using MinMaxScaler in the scikit-learn package~\cite{sklearn} so that it ranges from $-\pi$ to $\pi$.
The events are then passed to the machine learning algorithms, whether classical or quantum.

\begin{figure}[!htb]
   %\begin{minipage}{0.48\textwidth}
    \centering
    \includegraphics[scale=0.4]{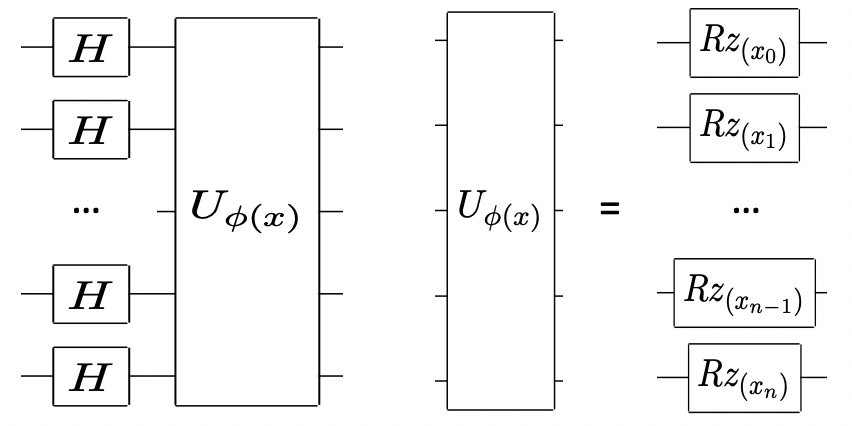}\\
    \textbf{(a)}\\
    \vspace{6pt}      
    \includegraphics[scale=0.4]{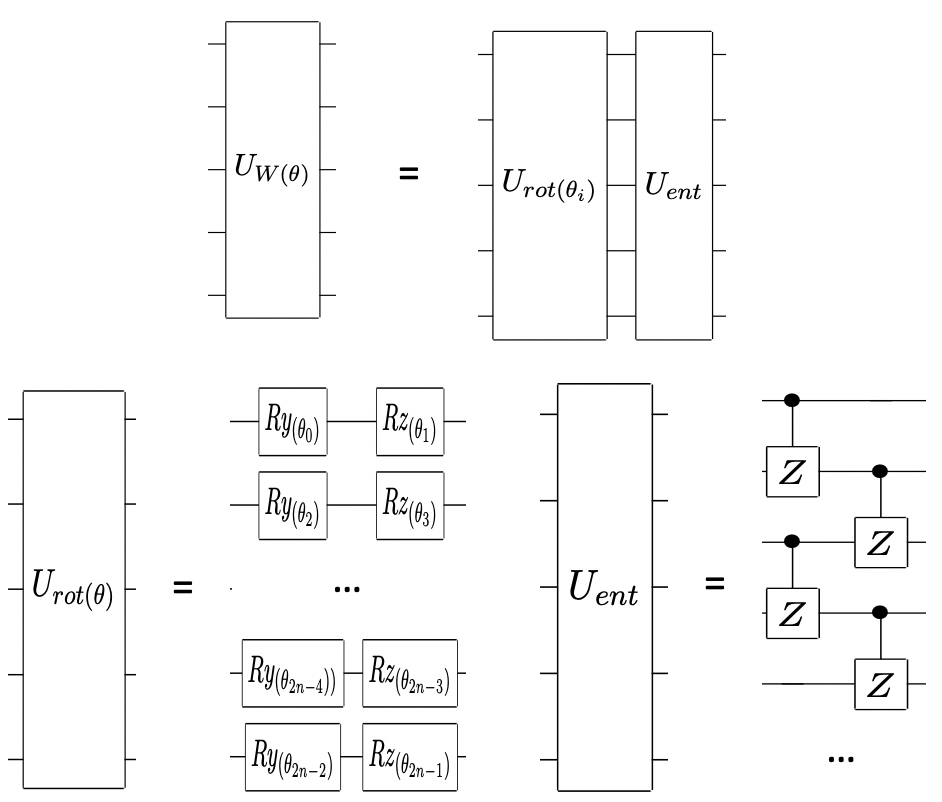}\\
    \textbf{(b)}\\
    \vspace{6pt}      
    \includegraphics[scale=0.6]{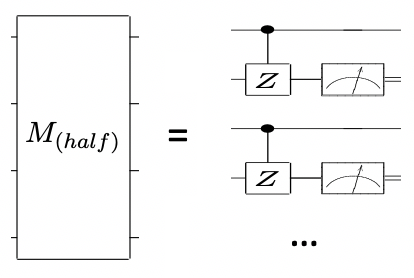}\\
    \textbf{(c)}\\
    \caption{
    Quantum circuits used in our quantum variational classifier studies.
    \textbf{(a)} The quantum feature map. First, Hadamard gates $H$ initialize every qubit to an equal superposition of all basis states. Then the feature map circuit $U_{\Phi(\vec{x})}$ encodes classical variables to quantum states by applying a phase rotation $RZ$ of an angle $x_i$. 
    The parameterization of $x_i$ with the corresponding PCA variable is slightly different between the \ttH\ and the \Hmumu\ cases. The feature map circuits can be duplicated multiple times with a depth parameter.
    \textbf{(b)} The quantum variational circuits $W(\vec{\theta})$. The variational rotation circuit $U_{\textrm{rot}}(\theta)$ is parameterized by ${\theta}$ and is followed by the entanglement circuit $U_{ent}$. $U_{\textrm{rot}}(\theta)$ consists of two rotations, $RY$ and $RZ$, on every qubit with parameterized ${\theta}$. For the entanglement step $U_{ent}$, we use the controlled phase gate $CZ$ to entangle adjacent qubits. In order to parallelize the qubit operations, we optimized the $U_{ent}$ circuit to reduce  execution dependency by at first only entangling each even qubit to its following odd qubit and then only entangling each odd qubit to its following even qubit.
    Multiple copies of the variational and entanglement circuits with another depth parameter can be applied to increase the number of degrees of freedom in a machine learning model~\cite{qsvmv}.
    \textbf{(c)} The measurement circuit $M_{(half)}$. To measure half of the qubits, a controlled phase gate operation $CZ$ is applied on every two qubits and only one of the two entangled qubits is measured.}

   \label{fig_qfmap_qvar}
   %\end{minipage}\hfill
   %\begin{minipage}{0.48\textwidth}
   %\end{minipage}
\end{figure}

%\section{Quantum variational algorithm and workflow}
\clearpage
\textbf{Quantum variational classifier algorithm and workflow:}
Following~\cite{qsvmv}, we use the quantum variational classifier algorithm to classify physics events of interest from background events. 
%This quantum approach exploits the mapping of the input data to an exponentially large quantum state space to enhance the ability to find an optimal classification solution.
This quantum approach exploits the mapping of classical input data to an exponentially large quantum feature space, which is based on quantum circuits that are hard to simulate classically.
It can be summarized in four main steps:
\begin{enumerate}
    \item Apply a feature map circuit $U_{\Phi(\vec{x})}$ to encode the input data $\vec{x}$ (containing 10 PCA variables) into a quantum state $\ket{\Phi(\vec{x})}$, as shown in Figure~\ref{fig_qfmap_qvar}(a). In our study the feature map encodes N classical variables to the quantum state space of a N-qubit system. 
    
    \item Apply a quantum variational circuit $W(\vec{\theta})$ parameterized by gate angles $\vec{\theta}$, as shown in Figure~\ref{fig_qfmap_qvar}(b). Here the variational circuit takes the form \begin{equation}
        W(\vec{\theta}) = U_{\textrm{rot}}(\theta) \;U_{\textrm{ent}} \ldots U_{\textrm{rot}}(\theta) \; U_{\textrm{ent}}      \end{equation}
    where $U_{\textrm{rot}}(\theta)$ refers to a variational circuit consisting of rotations on different qubits and $U_{ent}$ refers to entanglement unitary operations~\cite{qsvmv}.
    
    \item Measure the qubit state in the computational basis.
    The qubit measurement error is one of the largest error sources on the quantum hardware and results in imprecision on the classification result. To reduce this imprecision, we entangle every two qubits and then measure half of the N qubits, as shown in Figure~\ref{fig_qfmap_qvar}(c). 
    %As the hardware has a limitation on the number of shots, there will be lack of enough statistics if measuring too many qubits, especially for some low probability quantum states. 
    %Even more, the impact of noise and measurement errors will aggravate this imprecision.
    %To make sure that there are enough statistics on different quantum states and to reduce the impact of measurement errors on the statistics of different quantum states, 
    %To reduce this imprecision, an entanglement operation $M_{(half)}$ is applied to entangle every two qubits and then only one qubit is measured, as shown in Figure~\ref{fig_qfmap_qvar}(c). 
    %In fact, different entanglement solutions can be applied to measure more qubits or less qubits. Some experiments only measure one qubit, for example ~\cite{qneuron}.
    Additionally, a measurement error mitigation method implemented by the Qiskit framework~\cite{qiskit} is applied when measuring qubits. This error mitigation method 
    %first generates a simple circuit with known ideal results. These are compared with results from the actual execution of this circuit, which includes noise. 
    derives a relation matrix between the ideal results and the noisy results, which is later used to correct the noisy results.

    \item Classify the state through the action of a diagonal operator $f$ in computational basis with the eigenvalue being either $+1$ or $-1$. A discriminant is evaluated for the input data $\vec{x}$ according to
    \begin{equation}
    %p_y(\vec{x}) \leftarrow \bra{\Phi(\vec{x})} W^\dag(\vec{\theta}) M_y W(\vec{\theta}) \ket{\Phi(\vec{x})}~\cite{Havlcek2019SupervisedLW}
    \langle \Phi(\vec{x}) | W^\dag(\vec{\theta}) 2^{-1} (1+f) W(\vec{\theta}) | \Phi(\vec{x}) \rangle      
    \end{equation}
    and used to assign an output label $y \in \{1, 0\}$ denoting either a signal or background process~\cite{qsvmv}.
\end{enumerate}

%Following~\cite{qsvmv} we propose to look for new mapping of the classical SVM approach into a quantum algorithm in which the feature map is evaluated in the Hilbert space of the N-qubit system. 
%Following~\cite{qsvmv}, we look into the quantum variational algorithm. 
%This approach can be summarized in four main steps: i) loading of the data using a quantum feature map $\Phi$(x) that encodes N classical variables to the quantum state space of a N-qubit system, ii) evaluation of a short-depth circuit W($\vec{\theta}$) parameretized in the gate angles $\vec{\theta}$, iii) measurement of the qubit state z in the computational basis, and iv) the labeling (classification) of the state through the action of a switching function f(z). Both the feature map $\Phi$(x) and variational circuit W($\vec{\theta}$) provide entanglements over the qubits. 

During the training phase, a set of input data $\vec{x}$ and corresponding outputs $y$ are used to train the circuit W($\vec{\theta}$) to reproduce the correct classification. 
%To this end, an empirical cost function is defined by the error probability of incorrect assignment compared to the exact solutions available for the training set. 
%The training process consists in learning $\vec{\theta}$ to minimize the loss between the predicted $p_y(\vec{x})$ and the known classification label $y$. 
%Different optimizers, such as COBYLA~\cite{COBYLA} and SPSA~\cite{spsa1, spsa2}, can be applied.
The set of optimized parameters $\vec{\theta}$ is then kept fixed for all future classifications of the physical data. 

\clearpage

%\section{Result from the IBM Quantum Simulator with 10 qubits}

\textbf{Result from the IBM Quantum Computer Simulator with 10 qubits:}
We employ quantum machine learning with 10 qubits on the \textit{ibmq QasmSimulator}~\cite{qiskit} to classify signal and background processes for the \ttH\ analysis and the \Hmumu\ analysis. 
The \textit{ibmq QasmSimulator} simulates executions and measurements on quantum circuits of the IBM quantum computer hardware. 
The simulation incorporates a noise model generated from the properties of real hardware device. 
%although the qubit relaxation time (T1) and dephasing time (T2) are currently neglected.
In each analysis, we apply the quantum variational classifier algorithm to ten independent datasets, each consisting of 100 events for training and 100 events for testing. 
The quantum circuits are optimized to best fit the constraints imposed by the hardware (e.g., qubit connectivity, gate set availability, and hardware noise), as well as the nature of the data.
In the optimized configuration, the feature map depth is 1 and the variational circuit depth is 1. 
%and the maximum number of CNOT (Controlled-Not) gates in an execution path is 14.
With the present status of our accessed  hardware, the limited circuit depths are adopted to overcome the
hardware noise when utilizing 10 qubits. 
The circuit implementation uses linear qubit connectivity. 
The Spall's simultaneous perturbation stochastic approximation (SPSA) algorithm~\cite{spsa1,spsa2} is used as the optimizer for the variational circuit parameters $\vec{\theta}$ in the training process.
%because it performs relatively well in a noisy environment. 
%The qubit measurement error is one of the largest error sources on the quantum hardware. 
%In order to reduce its impact, we measure 5 qubits rather than all of the 10 qubits, 
%and entangle the other 5 qubits to the 5 measured qubits.
%Additionally, we apply error mitigation to the measured qubits.
With the same ten datasets and the same 10 variables processed with the PCA method, we also train a classical SVM~\cite{svm} classifier using the scikit-learn package~\cite{sklearn} and a BDT~\cite{bdt1,bdt2} classifier using the XGBoost package~\cite{xgboost}. 
The classical SVM and the BDT serve as benchmarks for classical machine learning algorithms.
Hyper-parameter tuning was performed on these classical algorithms.

\begin{figure}[htb]
\begin{center}
\includegraphics[width=5.0in]{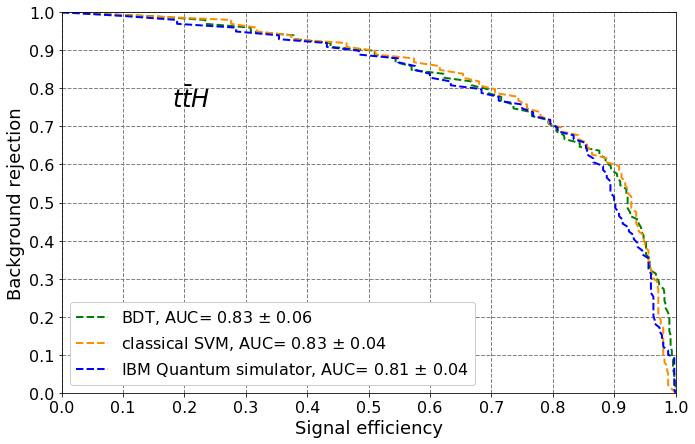}\\
\textbf{(a)} \\
\vspace{6pt}
\includegraphics[width=5.0in]{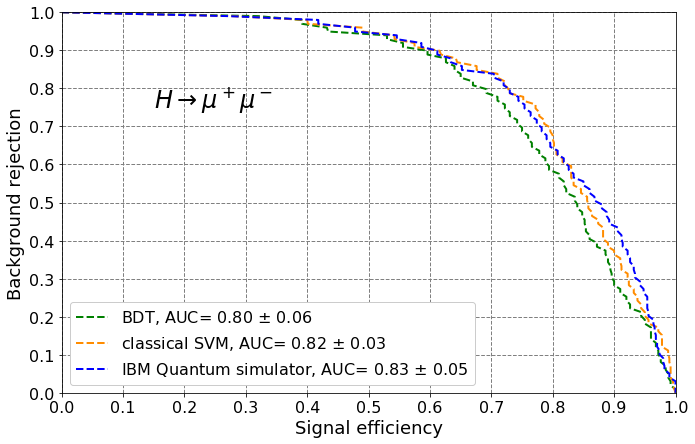}\\
\textbf{(b)}\\
\end{center}
\caption{The Receiver Operating Characteristic (ROC) curves (as a benchmark in the plane of background rejection versus signal efficiency) of the quantum variational classifier method on the \textit{ibmq QasmSimulator} (blue), the classical SVM (yellow), and the BDT (green) for (a) the \ttH\ analysis and (b) the \Hmumu\ analysis.
In each analysis, the classifiers are constructed using ten independent datasets, each consisting of 100 events for training and 100 events for testing. 
All classifiers are trained with the same 10 variables processed with the PCA method. 
In this study, 10 qubits are employed on the quantum computer simulator. 
%We have finished 1000 optimization iterations on the Quantum Computer Hardware and on the Quantum Computer Simulators. 
To visualize the discrimination power of each algorithm, the testing events of the ten datasets are combined to make the ROC curves.
We observe that the quantum variational classifier method on the \textit{ibmq QasmSimulator} performs similarly to the classical SVM and the BDT for both the \ttH\ analysis and the \Hmumu\ analysis. %The discrimination power provided by the Quantum Computer Hardware (red curve) is currently about 10$\%$ below the Quantum Computer Simulators (blue curve) and the classical machine learning algorithms (BDT: green curve; classical SVM: yellow curve).
}
\label{fig2}
\end{figure}

To study the discrimination power of each algorithm for both \ttH\ and \Hmumu, the testing events of the ten datasets are combined to make Receiver Operating Characteristic (ROC) curves as a benchmark in the plane of background rejection versus signal efficiency, as shown in Figure~\ref{fig2}.
%The ROC curves of the quantum variational method using the quantum simulator (blue curves), the classical SVM (yellow curves) and the BDT (green curves) appear to be similar.
We observe that in both the \ttH\ analysis and the \Hmumu\ analysis,  the quantum variational classifier method on the \textit{ibmq QasmSimulator} (blue) performs similarly to the classical SVM (yellow) and the BDT (green). 
%We observe that the quantum variational method on the QASM quantum simulator (blue) performs similarly to the classical SVM (yellow) and the BDT (green). 
We quantify the discrimination power of each classifier by the AUC (area under the ROC curve). 
In the \ttH\ analysis, the AUC for the quantum variational classifier method reaches 0.81$\pm$0.04 on the \textit{ibmq QasmSimulator}, 
compared to 0.83$\pm$0.04 for the classical SVM and 0.83$\pm$0.06 for the BDT.
Similarly, in the \Hmumu\ analysis, the AUC for the quantum variational classifier method reaches 0.83$\pm$0.05 on the \textit{ibmq QasmSimulator}, 
compared to 0.82$\pm$0.03 for the classical SVM and 0.80$\pm$0.06 for the BDT.
The quoted errors are the standard deviations for the AUC values of the ten datasets.
%This demonstrates the capability of quantum machine learning in distinguishing signal from background on realistic physics datasets. \\
This demonstrates the quantum algorithm can accurately distinguish signal from background on realistic physics datasets; the performance is comparable to (within the margin of error) state-of-the-art classical methods.\\

Ultimately, we can select events based on the classifier discriminant to maximize the quantity $S/\sqrt(B)$, where $S$ is the number of signal events and $B$ is the number of background events remaining after the selection. $S/\sqrt(B)$ is an approximation of the signal significance and is typically correlated with the classifier AUC, as both indicate the degree of separation between signal and background.
In the $\ttH$ analysis, a selection on the variational quantum classifier discriminant with a signal acceptance of 0.70 is associated with a background rejection of 0.78 (see the ROC curve in Figure~$\ref{fig2}$ (a)), and hence improves $S/\sqrt(B)$ by approximately $0.70/\sqrt(0.22)-1=50\%$ with respect to no selection.
Similarly, in the $\Hmumu$ analysis, a selection on the variational quantum classifier discriminant with a signal acceptance of 0.70 is associated with a background rejection of 0.84 (see the ROC curve in Figure~$\ref{fig2}$ (b)), hence improving $S/\sqrt(B)$ by approximately $0.70/\sqrt(0.16)-1=75\%$.

%figure 2

%We also perform a preliminary study applying the quantum SVM machine learning algorithm to the Higgs to two muons analysis with Delphes simulation events using the IBM quantum computer simulators.\\

\clearpage

%\section{Result from the IBM Quantum Hardware with 10 qubits} 

\textbf{Result from the IBM Quantum Computer Hardware with 10 qubits:}
At this point, it is interesting to assess the potential of quantum hardware calculations on the classification of the data presented in the previous section, and quantify the effect of the device noise. 
For the \ttH\ analysis and the \Hmumu\ analysis, we employ the quantum variational classifier algorithm with 10 qubits on the \textit{``ibmq\_boeblingen''} and \textit{``ibmq\_paris''} quantum computer hardware. 
%Boeblingen and Paris are two 20-qubit quantum processors based on superconducting electronic circuits. 
\textit{``ibmq\_boeblingen''} is a 20-qubit quantum processor and \textit{``ibmq\_paris''} is a 27-qubit quantum processor. Both are based on superconducting electronic circuits. 
%On the quantum hardware, the quantum circuit optimization is guided
%by the constrained imposed by the hardware (e.g., qubit connectivity, gates set availability, and affordable circuit depth) and by the nature of the data.
%In the optimized configuration, the feature map depth is 2, the variational circuit depth is 3, and the CNOT (Controlled-Not) gate depth is 14.
%Additionally, we apply error mitigation to the measurement qubits.
Due to current limitation of the access time to the quantum processors, %on Boeblingen, 
the quantum variational classifier algorithm is only applied to one of the ten datasets for each physics analysis.
%We pick the dataset whose ROC curve AUC on the QASM quantum simulator is the most close to the average of the ten datasets. 
We pick the dataset whose simulator AUC is closest to the average simulator AUC of the ten datasets.
The circuit, optimizer, and error mitigation configuration on the hardware is kept the same as for the simulator jobs. 
%We have finished 1000 iterations on the hardware and on the simulators. (An iteration is a term that indicates the number of times the algorithm's trainable parameters are updated.)

\begin{figure}[htb]
\begin{center}
\includegraphics[width=4.6in]{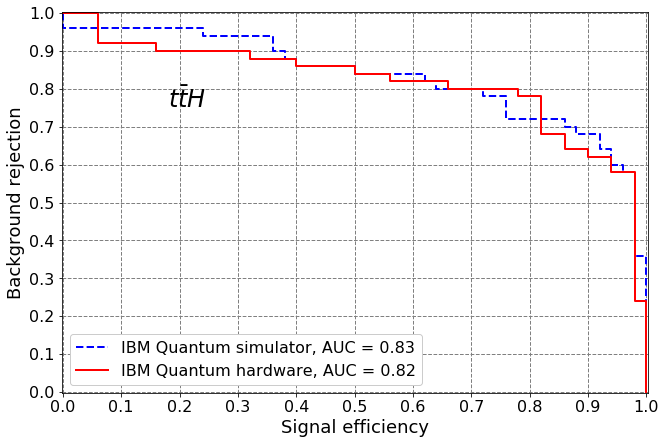}\\
\textbf{(a)} \\
\vspace{6pt}
\includegraphics[width=4.6in]{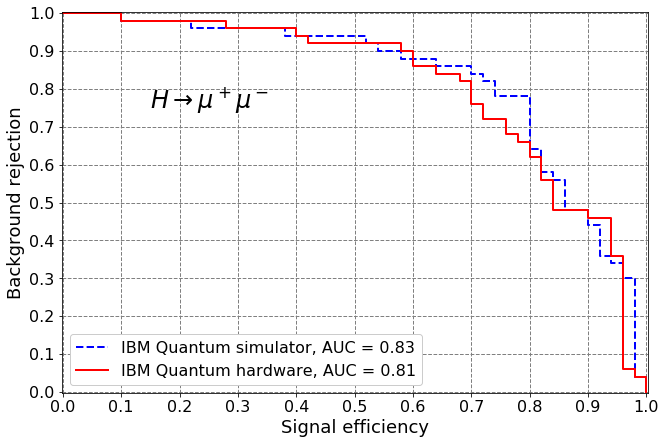}\\
\textbf{(b)}\\
\end{center}
\caption{
The Receiver Operating Characteristic (ROC) curves of the quantum variational classifier method with the \textit{``ibmq\_boeblingen''} and \textit{``ibmq\_paris''} quantum computer hardware (red) and with the \textit{ibmq QasmSimulator} (blue) for (a) the \ttH\ analysis (using \textit{``ibmq\_boeblingen''}) and (b) the \Hmumu\ analysis (using \textit{``ibmq\_paris''}).
%Results of the classical SVM (yellow curves) and the BDT (green curves) are overlaid. 
For each physics analysis, one dataset consisting of 100 events for training and 100 events for testing is utilized to construct the classifiers. 
This dataset is one of the ten datasets used in Figure~\ref{fig2}.
All classifiers are trained with the same 10 variables processed with the PCA method. 
In this study, 10 qubits are employed on the quantum computer hardware and the quantum computer simulator. 
%We have finished 1000 optimization iterations on the Quantum Computer Hardware and on the Quantum Computer Simulators. 
To visualize the discrimination power of both the quantum simulator and quantum hardware, the testing events of the dataset are used to make the ROC curves.
We observe that, for the quantum variational classifier method, 
%the discrimination power on the quantum hardware is close to that on the quantum simulator.
the quantum simulator and quantum hardware results appear to be in good agreement. 
%In our attempts, 
%With the circuit configuration optimized for 10 qubits, 
%the qubit platform quantum computers have achieved reasonable performance in exploiting a quantum state space with $2^{10}$ dimensions to distinguish different physics processes at the LHC.
%We observe the Quantum Computer Simulator and the BDT show similar performance, while the Quantum Computer Hardware currently gives worse discrimination power. 
}
\label{fig3}
\end{figure}

The ROC curves of the quantum variational classifier algorithm on the \textit{`ibmq\_boeblingen''} quantum hardware (for \ttH) and \textit{``ibmq\_paris''} quantum hardware (for \Hmumu) are shown in red in Figure~\ref{fig3}.
The ROC curves for the \textit{ibmq QasmSimulator} with the same datasets are overlaid in blue.
We observe that for the quantum variational classifier method, 
%the discrimination power on the quantum hardware is close to that on the quantum simulator.
the quantum simulator and quantum hardware results appear to be in good agreement. 
In the \ttH\ analysis, the quantum hardware AUC is 0.82, while the quantum simulator AUC is 0.83.
Similarly, in the \Hmumu\ analysis, the quantum hardware AUC is 0.81, while the quantum simulator AUC is 0.83.
In each analysis, the difference between the hardware AUC and the simulator AUC is found to be compatible with the test sample statistical error evaluated using a Bootstrapping re-sampling method.
%In our attempts, 
With the circuit configuration optimized for 10 qubits, 
the gate-model quantum computers have achieved reasonable performance in exploiting a quantum state space with $2^{10}$ dimensions to distinguish signal from background at the LHC.
%The potential difference could be due to the effect of the quantum hardware noise. \\
%Increasing the number of iterations is also expected to improve the performance. 
%(We continue to increase gradually the number of iterations to 1000.)\\

%figure 5

%figure 4
\begin{figure}[htb]
\begin{center}
\includegraphics[width=4.8in]{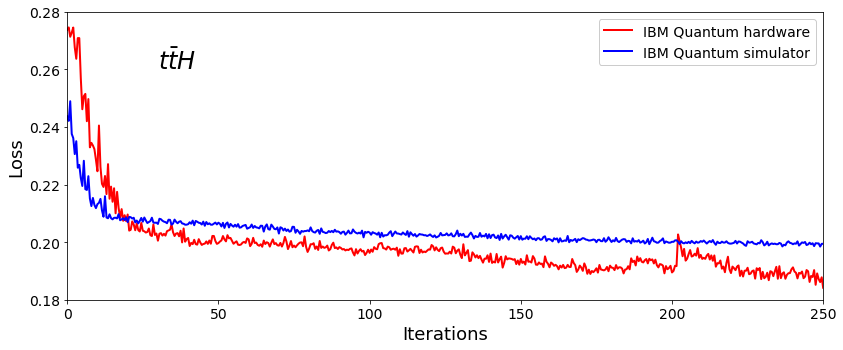}\\
\textbf{(a)}\\
\vspace{6pt}
\includegraphics[width=4.8in]{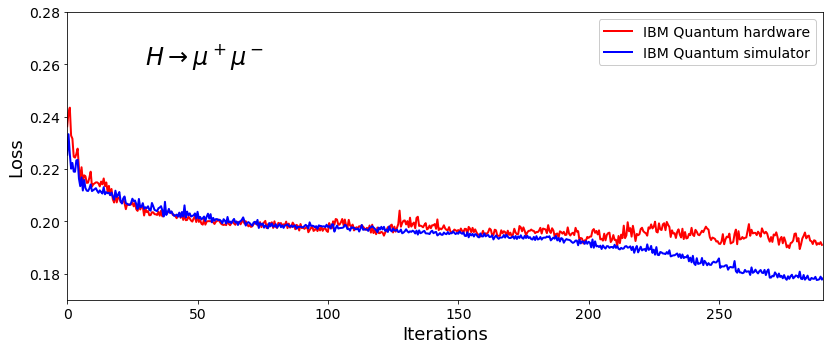}\\
\textbf{(b)}\\
\end{center}
\caption{The evolution of the loss function versus the number of iterations during the training process of the quantum variational classifier on the quantum computer hardware (red) and quantum computer simulator (blue) for (a) the \ttH\ analysis and (b) the \Hmumu\ analysis.
The number of iterations indicates the number of times the variational circuit parameters are updated in the training process.
The empirical loss function is defined by the error probability of incorrect assignment compared to the exact solutions available for the training set. 
%On the Quantum Computer Hardware, We have finished 1000 optimization iterations. 
%The loss function is the mean of the squared differences between the predicted and actual values of event classes. 
%The accuracy is the ratio of the number of correct predictions to the total number of predictions.
%The loss function, the accuracy and the ROC curve AUC are widely used to evaluate how well a machine learning algorithm performs.
The loss function improves and converges as the number of iterations increases, indicating that the quantum algorithm on hardware is indeed learning the difference between signal and background.
%It indicates that, the qubit platform quantum computers have the ability to learn the difference between the signal and the background in a high energy physics analysis.
}
\label{fig4}
\end{figure}

Figure~\ref{fig4} shows the evolution of the loss function versus the number of iterations during the training process of the quantum variational classifier on the hardware (red) and simulator (blue) for the \ttH\ analysis and the \Hmumu\ analysis.
The number of iterations indicates the number of times the variational circuit parameters are updated in the training process.
The empirical loss function is defined by the error probability of incorrect assignment compared to the exact solutions available for the training set. 
%For each event, the quantum variational algorithm outputs a score representing the probability for the event to be from the signal process. The score is then used for event class prediction. 
%In the ideal case, the scores of all signal events will be 1 and the scores of all background events will be 0. 
%The loss function is defined as the mean of the squared differences between the output scores from the quantum algorithm and the ideal scores. 
During the training process, the loss function is minimized to penalize misassignment and to optimize classifier parameters.
The loss function improves and converges as the number of iterations increases, indicating that the quantum algorithm on hardware is indeed learning the difference between signal and background in a realistic high energy physics analysis at the LHC.
%The performance of the QSVM training process is stable after 750 iterations.

%In our study, the running time of 500 training iterations on quantum hardware is 202 hours for 100 events of the \ttH\ analysis and 203 hours for 100 events of the \Hmumu\ analysis. The running time is currently longer than the classical algorithms, because today’s quantum hardware is not yet fully mature. With more mature quantum hardware in the future, we expect to see quantum speed-ups in quantum machine learning applications to high energy physics. 

In our study, 200 hours are required to run 500 training iterations on 100 events of the $\ttH$ or $\Hmumu$ analyses on quantum hardware. This is longer than that of the classical algorithms because today's quantum hardware is not yet fully mature. However, with the present rapid developments in quantum hardware, we expect in the future to see speed ups in quantum machine learning applications to high energy physics. 

%\begin{itemize}
%\item Collaborating with IBM Research, Zurich, we performed training with machine learning on the IBM Quantum Computer Hardware with 100 training events, 100 test events, and 5 qubits, again for the ttH (H to two photons) analysis at the LHC. Because of hardware access time and timeout limitations, we only finished a few iterations (for example 10, 30, 50) on the hardware, compared to several thousand iterations on the simulators. The limitation of access time for 5 qubits or higher (10, 20 qubits) will be alleviated in our future collaboration with IBM.

%\item As shown in Figure~\ref{fig6}, with limited iterations, the result from hardware (blue square) is compatible with the result from the Quantum Simulator (red) in tested iterations. The result from the Quantum Simulator (red) reached a similar performance to that of the classical BDT method (green) with enough iterations. (this bullet is wrong)
%\end{itemize}

\clearpage

%\section{SUMMARY}
%\vspace{0.2cm}

\textbf{Conclusion:}
%The era of efficient Quantum Computing may still be some years away. 
In this study, we have obtained early results in the application of quantum machine learning with 10 qubits 
on the \textit{imbq QasmSimulator} and the \textit{``ibmq\_boeblingen''} and \textit{``ibmq\_paris''} quantum hardware to two recent LHC flagship physics analyses: \ttH\ and \Hmumu.
\ttH, Higgs boson production in association with a top quark pair, probes the Higgs boson couplings to the top quark, while \Hmumu, Higgs boson decays to two muons, probes the Higgs boson couplings to second-generation fermions. 
In this study we do not attempt to do a  complete analysis of $\ttH$ and $\Hmumu$. Rather our goal is to perform proof of principle in using quantum machine learning compared with popular classical machine learning methods, BDT for example.
With small training samples of 100 events, the quantum variational classifier method on the \textit{ibmq QasmSimulator} performs similarly to the classical SVM algorithm and the BDT algorithm. 
%demonstrating the feasibility of this quantum machine learning algorithm.
The quantum variational classifier method on the quantum hardware has shown promising discrimination power comparable to that on the quantum simulation. 

To study the discrimination power of quantum machine learning classifiers, we make use of Receiver Operating Characteristic (ROC) curves in the plane of background rejection versus signal efficiency as a standard metric in machine learning application for the high energy physics.
We further quantify the discrimination power of the classifiers by the AUC (area under the ROC curve). 
The use of ROC curves and AUCs to be the metric of discrimination power compared with classical machine learning methods is inspired by Ref~$\cite{maria}$. A difference is that Ref~$\cite{maria}$ uses quantum annealers while our work uses gate-based quantum computers.
In the \ttH\ analysis, the quantum hardware AUC is 0.82, while the quantum simulator AUC is 0.83.
Similarly, in the \Hmumu\ analysis, the quantum hardware AUC is 0.81, while the quantum simulator AUC is 0.83.
These results demonstrate that quantum machine learning on the hardware of the gate-model quantum computers has the ability to differentiate between signal and background in a realistic high energy physics analysis at the LHC.
%\hl{Furthermore, although we have demonstrated that the quantum and classical machine learning algorithms perform similarly, with the rapid advance of the quantum computing technology, the quantum computers may offer speedups in machine learning}~\cite{qml}, \hl{which can be critical for the HEP community in the future.}
Furthermore, although we have demonstrated that the quantum and
classical machine learning algorithms perform similarly, with the rapid
advance of the quantum computing technology, the use of quantum machine
learning may offer a ``speed up'' advantage~\cite{qml}, which can be critical for the future of the high energy physics community.

%The quantum hardware performance could be further improved by reducing the impact of quantum hardware noise. 
%and increasing the number of iterations on the hardware. 
%For the \Hmumu\ analysis, the Quantum Hardware and the Quantum Simulators show similar performance.

In the future, by exploiting the high dimensional feature space defined by a larger number of qubits and by mitigating the impact of quantum hardware noise, quantum machine learning classifiers could possibly outperform classical classifiers.
We plan to explore quantum algorithms to extend our analysis to more qubits and larger sample sizes. 
Moreover, we plan to apply the quantum kernel classifier method proposed in~\cite{qsvmv} to our LHC flagship physics analyses.
%Additionally, we will work on error mitigation in the context of using noisy intermediate-scale quantum (NISQ) hardware~\cite{nisq}. 
We foresee the usage of quantum machine learning in future high-luminosity LHC physics analyses, including measurements of the Higgs boson self-couplings and searches for dark matter. 
\\

%Furthermore, since quantum hardware is still in an intensive state of development, we need to work out a more effective way to apply the quantum simulator to high energy physics, given the importance of simulation power in this early work before we have quantum hardware with a larger number of qubits.\\

\clearpage

\vspace{0.5cm}
\textbf{Acknowledgements } 
This project is supported in part by the United States Department of Energy, Office of Science, HEP-QIS Research Program, under Award Number DE-SC0020416.
This project is supported in part by the United States Department of Energy, Office of Science, Office of High Energy Physics program under Award Number DE-SC-0012704 and the Brookhaven National Laboratory LDRD \#20-024.
This research used resources of the Oak Ridge Leadership Computing Facility, which is a United States Department of Energy Office of Science User Facility supported under Contract DE-AC05-00OR22725.
The Wisconsin group would like to thank the ATLAS Collaboration for the inspiration of the two LHC flagship analyses used in this
publication.

%\vspace{0.5cm}
%\textbf{Author contributions }
%S.L.W. conceived this study in application of quantum machine learning on the recent LHC flagship analyses, 
%assembled the collaboration, and together with F.C. and A.D.M. oversaw the progress of this project.
%W.G. and S.S. adapted and debugged the algorithms and workflows for the application of the quantum variational classifier method to the LHC flagship analyses, and performed quantum simulation. 
%W.G., S.S, A.C.Y.L., J.L. and P.S. optimized circuits and configurations on the IBM quantum hardware.
%A.C.Y.L., J.L., P.S., S.Y., S.Y.-C.C. and T.-C.W., submitted and monitored our recent LHC flagship analyses on the IBM quantum hardware.
%C.Z., A.W. and J.C. provided simulated event samples for the LHC flagship analyses, and compared the classical machine learning methods with the quantum variational classifier method.
%S.L.W. and C.Z. interpreted and critically evaluated the final results and wrote the initial manuscript, 
%while W.G. extensively reviewed and revised the manuscript.
%M.L., F.C. and A.D.M. solved the challenge of heavy computing resources and large memory necessary for quantum simulation. 
%P.B., I.T., S.W. and J.G. provided quantum information science expertise to solve the quantum simulator and hardware challenges.
%A.C.Y.L., J.L., P.S., S.Y.-C.C., S.Y. and T.-C.W. submitted and monitored our LHC flagship analyses on the IBM quantum hardware.
%Assisted by W.G., S.L.W. and C.Z. contributed to writing the manuscript.
%All authors contributed to the writing and reviewing of the manuscript.

\vspace{0.5cm}
\textbf{Correspondence and requests for materials} should be addressed to S.L.W. (Sau.Lan.Wu@cern.ch).

%\vspace{0.5cm}
%\textbf{Data Availability } The data of this study are available to any interested researchers on request. 

%\vspace{0.5cm}
%\textbf{Additional Information} The authors declare no competing interests.


\clearpage

\begin{thebibliography}{99}

\bibitem{atlashiggs} ATLAS Collaboration, Aad, G. et al. Observation of a new particle in the search for the standard model Higgs boson with the ATLAS detector at the LHC. \emph{Phys. Lett. B} \textbf{716}, 1–29 (2012).

\bibitem{cmshiggs} CMS Collaboration, Chatrchyan, S. et al. Observation of a new boson at a mass of 125 GeV with the CMS experiment at the LHC. \emph{Phys. Lett. B} \textbf{716}, 30–61 (2012).

%\bibitem{mlopt_quant} Dasari, V. R., Im, M. S., Beshaj, L., \emph{Solving machine learning optimization problems using quantum computers}, \emph{arXiv:1911.08587}

%\bibitem{qml_cls} Ciliberto, C., Herbster, M., Ialongo, A. D., Pontil, M., , A., Severini S., Wossnig L., \emph{Quantum machine learning: a classical perspective}, \emph{Proc. R. Soc. A}, vol. 474, no. 2209, p. 20170551. The Royal Society, (2018).

%\bibitem{nisq} Preskill, J. Quantum Computing in the NISQ era and beyond. \emph{Quantum} \textbf{2}, 79 (2018).

\bibitem{qml} Biamonte, J., Wittek, P., Pancotti, N., Rebentrost, P., Wiebe N., and Lloyd., S. Quantum machine learning. \emph{Nature} \textbf{549}, 195–202 (2017).

%\bibitem{qsvm} Rebentrost, P., Masoud, M. and Lloyd, S., \emph{Quantum support vector machine for big data classification}, \emph{Phys. Rev. Lett.} 113, 130503 (2014).

\bibitem{qsvmv} Havlíček, V., Córcoles, A. D., Temme, K., Harrow, A. W., Kandala, A., Chow, J. M. and Gambetta, J. M. Supervised learning with quantum-enhanced feature spaces. \emph{Nature} \textbf{567}, 209-212 (2019).

\bibitem{maria} Mott, A., Job, J., Vlimant, J., Lidar D., and Spiropulu, M., Solving a Higgs optimization problem with quantum annealing for machine learning. \emph{Nature} \textbf{550}, 375-379 (2017).

\bibitem{atlastth} ATLAS Collaboration, Aaboud, M. et al. Observation of Higgs boson production in association with a top quark pair at the LHC with the ATLAS detector. \emph{Phys. Lett. B} \textbf{784}, 173-191 (2018).

\bibitem{cmstth} CMS Collaboration, Sirunyan, A. M. et al. Observation of ttH Production. \emph{Phys. Rev. Lett.} \textbf{120}, 231801 (2018).

\bibitem{atlashmumu} ATLAS Collaboration, Aad, G. et al. A Search for the Dimuon Decay of the Standard Model Higgs Boson with the ATLAS Detector. \emph{Phys. Lett. B} \textbf{812}, 135980 (2021).

\bibitem{cmshmumu} CMS Collaboration, Sirunyan, A. M. et al. Evidence for Higgs boson decay to a pair of muons. \emph{JHEP} \textbf{01}, 148 (2021).

\bibitem{madgraph} Alwall, J., Frederix, R., Frixione, S., Hirschi, V., Maltoni, F., Mattelaer, O., Shao, H. -S., Stelzer, T., Torrielli, P. and Zaro, M. The automated computation of tree-level and next-to-leading order differential cross sections, and their matching to parton shower simulations. \emph{JHEP} \textbf{07}, 079 (2014).

\bibitem{pythia} Sjostrand, T., Mrenna, S. and Skands, P. Z. PYTHIA 6.4 physics and manual. \emph{JHEP} \textbf{05}, 026 (2006).

\bibitem{delphes} de Favereau, J., Delaere, C., Demin, P., Giammanco, A., Lema{\^i}tre, V., Mertens, A. and Selvaggi, M. DELPHES 3, a modular framework for fast simulation of a generic collider experiment. \emph{JHEP} \textbf{02}, 057 (2014).

\bibitem{pca} Pearson, K. On lines and planes of closest fit to systems of points in space. \emph{The London, Edinburgh, and Dublin Philosophical Magazine and Journal of Science}, \textbf{2}, 559-572 (1901).

\bibitem{pca1} Jolliffe, I. T. Principal component analysis, second edition. \emph{Springer Science+Business Media}, (2002).

\bibitem{sklearn} Pedregosa, F., Varoquaux, G., Gramfort, A., Michel, V., Thirion, B., Grisel, O., Blondel, M., Prettenhofer, P., Weiss, R., Dubourg, V., Vanderplas, J., Passos, A., Cournapeau, D., Brucher, M., Perrot, M., Duchesnay, E. Scikit-learn: machine learning in python. \emph{Journal of Machine Learning Research} \textbf{12}, 2825-2830 (2011).

%\bibitem{minmax} Patro S. G. K., Sahu K. K., Normalization: A Preprocessing Stage, \emph{arXiv}, 1503.06462, (2015).

%\bibitem{qneuron} Tacchino F., Macchiavello C., Gerace D., Bajoni D. An artificial neuron implemented on an actual quantum processor. \emph{npj Quantum Inf} \textbf{5}, 26 (2019).

%\bibitem{qsvmvpre} Guan, W., Sun, S. J., Wang, A., Wu, S. L., Zhou, C. and Carminati, F., \emph{Preliminary Development on HEP Data Analysis Using Quantum Computing Based on IBM Qiskit}, \emph{Quantum Computing for High Energy Physics workshop, CERN, Geneva, Switzerland}(2018).

%\bibitem{qsvmvcern} Chan, J., Guan, W., Sun, S. J., Wang, A., Wu, S. L., Zhou, C., Carminati, F., Barkoutsos, P., Tavernelli, I.,  Woerner,  S. and Zoufal,  C., \emph{Applying IBM Quantum Computing to LHC Physics Analysis of Higgs Coupling to Top Quarks}, \emph{CERN openlab Technical Workshop}, CERN, Geneva, Switzerland(2019).

%\bibitem{qsvmveqtc19} Chan,  J.,  Guan,  W.,  Sun,  S. J.,  Wang,  A.,  Wu,  S. L.,  Zhou,  C.,  Livny,  M.,  Meglio,  A. D.,  Carminati,  F., Barkoutsos, P., Tavernelli, I., Woerner, S. and Zoufal, C., \emph{Application of IBM Quantum Computing to LHC
%High Energy Physics Data Analysis}, \emph{European Quantum Technologies Conference (EQTC19) in Grenoble,
%France, the First International Conference of the QT Flagship} (2019).

%\bibitem{COBYLA} Powell, M. J. D., \emph{A Direct Search Optimization Method That Models the Objective and Constraint Functions by Linear Interpolation}, \emph{Mathematics and Its Applications} 275, (1994).

\bibitem{qiskit} Aleksandrowicz, G. et al. Qiskit: An Open-source Framework for Quantum Computing. \emph{https://qiskit.org} (2019).


\bibitem{spsa1} Spall, J. C. A one-measurement form of simultaneous perturbation stochastic approximation. \emph{Automatica} \textbf{33}, 109–112 (1997).

\bibitem{spsa2} Spall, J. C. Adaptive stochastic approximation by the simultaneous perturbation method. \emph{IEEE Trans. Automat. Contr.} \textbf{45}, 1839-1853 (2000).

\bibitem{svm} Boser, B. E., Guyon, I. M., Vapnik, V. N. A training algorithm for optimal margin classifiers. \emph{Proceedings of the 5th Annual ACM Workshop on Computational Learning Theory}, 144-152 (1992).

\bibitem{bdt1} Friedman, J. H. Stochastic gradient boosting. \emph{Computational Statistics $\&$ Data Analysis} \textbf{38}, 367-378
(2002).

\bibitem{bdt2} Hastie, T., Tibshirani, R., Friedman, J. The elements of statistical learning: data mining, inference, and prediction, second edition. \emph{Springer Science+Business Media}, (2009).

%\bibitem{bdt3} Quinlan, J. R. Induction of decision trees. \emph{Machine Learning} \textbf{1}, 81-106 (1986).

\bibitem{xgboost} Chen, T. and Guestrin, C. XGBoost: a scalable tree boosting system. \emph{Proceedings of the 22nd ACM SIGKDD International Conference on Knowledge Discovery and Data Mining}, 785-794 (2016).

%\bibitem{tthphyr} ATLAS Collaboration, \emph{Measurements of Higgs boson properties in the diphoton decay channel with 36 $fb^{-1}$ of pp collision data at $\sqrt{s}$ = 13 TeV with the ATLAS detector}, \emph{Phys. Rev. D} 98, 052005 (2018).

%\bibitem{tthconf} ATLAS Collaboration, \emph{Measurement of Higgs boson production in association with a $t\bar{t}$ pair in the diphoton decay channel using 139 $fb^{-1}$ of LHC data collected at $\sqrt{s}$ = 13 TeV by the ATLAS experiment}, ATLAS-CONF-2019-004 (2019).

%\bibitem{hmumuconf19} ATLAS Collaboration, \emph{A search for the dimuon decay of the Standard Model Higgs boson in pp collisions at $\sqrt{s}$ = 13 TeV with the ATLAS experiment}, ATLAS-CONF-2019-028 (2019).

%\bibitem{hmumuconf18} ATLAS Collaboration, \emph{A search for the rare decay of the Standard Model Higgs boson to dimuons in pp collisions at $\sqrt{s}$ = 13 TeV with the ATLAS experiment}, ATLAS-CONF-2018-026 (2018).






\end{thebibliography}
\end{document}